\documentclass[twocolumn,showpacs,amsfonts,aps,prc,floatfix,superscriptaddress]{revtex4-1}
\usepackage[colorlinks=true, pdfstartview=FitV, linkcolor=red, citecolor=blue, urlcolor=blue]{hyperref}
\usepackage{amsmath}
\usepackage{amssymb}
\usepackage{bm}
\usepackage{graphicx}
\usepackage{color}

\begin{document}

\title{
Landau and Eckart frames for relativistic fluids in nuclear collisions
}

\author{Akihiko Monnai}
\email[]{akihiko.monnai@kek.jp}
\affiliation{KEK Theory Center, Institute of Particle and Nuclear Studies, \\
High Energy Accelerator Research Organization (KEK),
1-1, Ooho, Tsukuba, Ibaraki 305-0801, Japan}
\date{\today}

\begin{abstract}
The quark matter created in relativistic nuclear collisions is interpreted as a nearly-perfect fluid. The recent efforts to explore its finite-density properties in the beam energy scan programs motivate one to revisit the issue of the local rest frame fixing in off-equilibrium hydrodynamics. I first investigate full second-order relativistic hydrodynamics in the Landau and the Eckart frames. 
Then numerical hydrodynamic simulations are performed to elucidate the effect of frame choice on flow observables in relativistic nuclear collisions. 
The results indicate that the flow can differ in the Landau and the Eckart frames but charged particle and net baryon rapidity distributions are mostly frame independent when off-equilibrium kinetic freeze-out is considered.
\end{abstract}

%\keywords{}

\maketitle

%%%%%%%%%%%%%%%%%%%%%%%%%%%%%%%%%%%%%%%%%%%%%%%%%%%%%%%%%%%%%%%%
\section{Introduction}
\label{sec1}
\vspace*{-2mm}
%%%%%%%%%%%%%%%%%%%%%%%%%%%%%%%%%%%%%%%%%%%%%%%%%%%%%%%%%%%%%%%%

The existence of the strongly-coupled quark-gluon plasma \cite{Yagi:2005yb} as a high-temperature phase of QCD has been partly motivated by a number of relativistic hydrodynamic analyses of high-energy nuclear collisions at BNL Relativistic Heavy Ion Collider \cite{Arsene:2004fa,Back:2004je,Adams:2005dq,Adcox:2004mh} and CERN Large Hadron Collider \cite{Aamodt:2010pa,ATLAS:2011ah,Chatrchyan:2012wg}. Modern versions of such analyses incorporate the effects of viscosity to take account of off-equilibrium processes in the system, which play important roles in quantitative understanding of the experimental data \cite{Wang:2016opj}. 

The theoretical framework of relativistic dissipative hydrodynamics, however, is still not completely understood, partially because one has to introduce relaxation effects to the off-equilibrium processes to avoid violating stability and causality \cite{Hiscock:1983zz,Hiscock:1985zz,Hiscock:1987zz}. Such extended frameworks are called the second-order theory \cite{Israel:1976tn,Israel:1979wp,Muronga:2001zk,Muronga:2003ta,Muronga:2006zw,Muronga:2006zx,Koide:2006ef,Tsumura:2006hn,Tsumura:2007wu,Tsumura:2009vm,Tsumura:2011cj,Tsumura:2012ss,Baier:2007ix,Romatschke:2009kr,Bhattacharyya:2008jc,Natsuume:2007ty,Lublinsky:2009kv,Betz:2009zz,Monnai:2010qp,Monnai:2018rgs,Molnar:2013lta,Denicol:2012cn,Denicol:2012vq,Denicol:2019iyh,PeraltaRamos:2009kg,Calzetta:2010au,Aguilar:2017ios,Florkowski:2015lra,Jaiswal:2015mxa,Tinti:2016bav,Harutyunyan:2018cmm,Mitra:2019jld} as opposed to the traditional linear response theory \cite{Landau,Eckart:1940te}, which is also known as the first-order theory, because the off-equilibrium correction of the respective order in terms of dissipative currents is taken into account in the entropy current of those theories.

Non-relativistic hydrodynamic flow can be defined as a local flux of particles. In relativistic systems, however, the definition of the flow is not trivial because the energy and the conserved number can flow separately in the presence of dissipative processes. 
There are two distinctive ways of defining the local rest frame for the flow: the Landau (or energy) frame \cite{Landau} and the Eckart (or conserved charge/particle) frame \cite{Eckart:1940te}. 
There have been decades of debate over the eligibility of the two definitions of the local rest frame
\cite{Tsumura:2006hn,Tsumura:2007wu,Tsumura:2009vm,Tsumura:2011cj,Tsumura:2012ss,Van:2007pw,Van:2011yn,Osada:2011gx,Minami:2012hs,Oliinychenko:2015lva,Sagaceta-Mejia:2018cao}. Most of the numerical analyses of hydrodynamic models for relativistic nuclear collisions so far do not give explicit consideration to the frame because the diffusion or the dissipation current is neglected, but the Landau frame is often considered to be a preferred choice when there is a theoretical need. This could be owing to the fact that the primary conserved charge in nuclear collisions is the net baryon number, which is often small at high energies; the Eckart frame cannot be defined when conserved charges are approximated to be negligible. There are several calculations \cite{Monnai:2012jc,Kapusta:2014dja,Denicol:2018wdp,Li:2018fow} that include the effects of baryon diffusion, which intrinsically implies that the Landau frame is chosen. 

The beam energy scan (BES) programs are being performed at RHIC. The exploration of mid-to-low beam energy regime is also planned at facilities including GSI Facility for Antiprotons and Ion Research (FAIR), JINR Nuclotron-based Ion Collider fAcility (NICA), CERN Super Proton Synchrotron (SPS), and JAEA/KEK Japan Proton Accelerator Research Complex (J-PARC) in order to elucidate the QCD phase structure at finite densities. It would be insightful to come back to the question of the flow frame in hydrodynamic models and investigate whether the choice of the frame can affect observables in those experiments.

In this paper, full second-order hydrodynamic equations are investigated in the Landau and the Eckart frames. Stability and causality conditions in the two frames are shown to be related to the correspondences between the first- and the second-order transport coefficients in those frames. Then the implications of a frame choice on the hydrodynamic evolution in heavy-ion systems are discussed focusing on the baryon diffusion and the energy dissipation. Numerical analyses are performed for rapidity distribution because the effects of the net baryon number would appear most prominently in the direction of the collision. Fragments of the shattered nuclei are the source of the conserved charge. 

The paper is organized as follows. Full second-order relativistic dissipative hydrodynamics is investigated in the Landau and the Eckart frames in Sec.~\ref{sec2}. Causality and stability conditions are discussed in Sec.~\ref{sec3}. Sec.~\ref{sec4} presents numerical demonstration of the effects of a frame choice in nuclear collisions. Discussion and conclusions are presented in Sec.~\ref{sec5}. The natural unit $c = \hbar = k_B = 1$ and the mostly-negative Minkowski metric $g^{\mu \nu} = \mathrm{diag}(+,-,-,-)$ is used in this paper.

%%%%%%%%%%%%%%%%%%%%%%%%%%%%%%%%%%%%%%%%%%%%%%%%%%%%%%%%%%%%%%%%
\section{Relativistic hydrodynamics in Landau and Eckart Frames}
\label{sec2}
\vspace*{-2mm}
%%%%%%%%%%%%%%%%%%%%%%%%%%%%%%%%%%%%%%%%%%%%%%%%%%%%%%%%%%%%%%%%

The ideal hydrodynamic flow is uniquely determined since the local fluxes of the energy and the charge densities are in the same direction, \textit{i.e.}, the directions of the eigenvector of the energy-momentum tensor and the conserved current match as $T^{\mu \nu} u_\nu = e u^\mu$ and $N^\mu = n u^\mu$. Here $e$ is the energy density and $n$ is the conserved charge density. On the other hand, the presence of the vector dissipative currents lead to the separation of the two local fluxes in relativistic systems. The Landau frame is chosen in the direction of the total energy flux so the dissipation of energy does not appear explicitly, $T^{\mu \nu} u^L_\nu = e_L u_L^\mu$. The Eckart frame is the choice of flow where the total conserved charge flux is diffusion-less $N^\mu = n_E u_E^\mu$. Here the subscripts $L$ and $E$ represent the Landau and the Eckart frames, respectively. The energy-momentum tensor, the conserved charge current, and the entropy current $s^\mu$ are assumed to be invariant under frame transformations \cite{Israel:1979wp}. 

The tensor decompositions read
\begin{eqnarray}
T^{\mu \nu} &=& e_L u_L^\mu u_L^\nu - (P_L + \Pi_L) \Delta_L^{\mu \nu} + \pi_L^{\mu \nu}, \\
N^\mu &=& n_L u_L^\mu + V_L^\mu ,
\end{eqnarray}
in the Landau frame and
\begin{eqnarray}
T^{\mu \nu} &=& e_E u_E^\mu u_E^\nu - (P_E + \Pi_E) \Delta_E^{\mu \nu} \nonumber \\
&+& W_E^\mu u_E^\nu + W_E^\nu u_E^\mu + \pi_E^{\mu \nu}, \\
N^\mu &=& n_E u_E^\mu ,
\end{eqnarray}
in the Eckart frame. Here $\Pi$ is the bulk pressure, $\pi^{\mu \nu}$ is the shear stress tensor, $W^\mu$ is the energy dissipation, $V^\mu$ is the baryon diffusion, and $\Delta^{\mu \nu} = g^{\mu \nu} - u^\mu u^\nu$ is the space-like projection. It can be immediately seen that the two frames become identical in the ideal hydrodynamic limit. The dissipative corrections to the energy and the conserved charge densities are neglected for simplicity \cite{Monnai:2018rgs}. Also I consider a system with a single charge conservation though the extension to general systems should be a straightforward task. 

In the following arguments, the vector dissipative currents $W_E^\mu$ and $V_L^\mu$ are considered and the shear and bulk viscous corrections are set aside for simplicity. When the dissipative corrections are much smaller than the equilibrium variables, the difference in the thermodynamic variables of the two frames $\Delta n_{E-L} = n_E - n_L$ and $\Delta e_{E-L} = e_E - e_L$ are, at a given space-time point,
\begin{eqnarray}
\Delta n_{E-L} &=& \frac{1}{2} \frac{V_L^\mu V^L_\mu}{n_L} + \mathcal{O}(\delta^3), \label{eq:dn}\\
\Delta e_{E-L} &=& - \frac{W_E^\mu W^E_\mu}{e_E + P_E} + \mathcal{O}(\delta^3), \label{eq:de}
\end{eqnarray} 
where the correction is of the second order in dissipative quantities. They indicate that the corrections to other thermodynamic variables, \textit{i.e.}, the pressure $P$, the entropy density $s$, the temperature $T$, and the chemical potential $\mu$ are of the second order. The corrections to the transport coefficients should also be of the second order because they are functions of the energy and the conserved charge densities. Hereafter the subscripts $L$ and $E$ are dropped for those variables for simplicity unless otherwise required. The flow difference $\Delta u^\mu_{E-L} = u_E^\mu - u_L^\mu$ is 
\begin{eqnarray}
\Delta u^\mu_{E-L} = \frac{V_L^\mu }{n} + \mathcal{O}(\delta^2) = - \frac{W_E^\mu}{e+P} + \mathcal{O}(\delta^2), 
\end{eqnarray} 
where the leading order correction is of the first order. 

The macroscopic variables are estimated using the conservation laws $\partial_\mu T^{\mu \nu} = 0$ and $\partial_\mu N^\mu = 0$, the equation of state $P=P(e,n_B)$, and the constitutive relations for the dissipative currents. 
In the Landau frame, the second-order causal expression of the baryon diffusion, based on an extended Israel-Stewart framework \cite{Israel:1976tn,Israel:1979wp,Monnai:2010qp}, reads
\begin{eqnarray}
V_L^\mu &=& \kappa_V \nabla_L^\mu \frac{\mu}{T} - \tau_V (\Delta_{L})^{\mu}_{\ \nu} D_L V_L^\nu \nonumber \\
&+& \chi_V^a V_L^\mu D_L \frac{\mu}{T} + \chi_V^b V_L^\mu D_L \frac{1}{T} + \chi_V^c V_L^\mu \nabla^L_\nu u_L^\nu \nonumber \\
&+& \chi_V^d V_L^\nu \nabla^L_\nu u_L^\mu + \chi_V^e V_L^\nu \nabla_L^\mu u^L_\mu, 
\label{eq:diffusion}
\end{eqnarray}
where $\kappa_V \geq 0$ is the baryon conductivity, $\tau_V \geq 0$ is the relaxation time for the baryon diffusion, and $\chi_V^{a,b,c,d,e}$ are second-order transport coefficients. $D = u^\mu \partial_\mu$ and $\nabla^\mu = \Delta^{\mu \nu} \partial_\nu$ are the time- and the space-like derivatives, respectively. Similarly, in the Eckart frame, the energy dissipation reads
\begin{eqnarray}
W_E^\mu &=& - \kappa_W \bigg( \nabla_E^\mu \frac{1}{T} + \frac{1}{T} D_E u_E^\mu \bigg) - \tau_W (\Delta_{E})^{\mu}_{\ \nu} D_E W_E^\nu \nonumber \\
&+& \chi_W^a W_E^\mu D_E \frac{\mu}{T} + \chi_W^b W_E^\mu D_E \frac{1}{T} + \chi_W^c W_E^\mu \nabla^E_\nu u_E^\nu  \nonumber \\
&+& \chi_W^d W_E^\nu \nabla^E_\nu u_E^\mu + \chi_W^e W_E^\nu \nabla_E^\mu u^E_\nu, 
\label{eq:dissipation}
\end{eqnarray}
where $\kappa_W \geq 0$ is the energy conductivity and $\tau_W \geq 0$ is the relaxation time for the energy dissipation, and $\chi_W^{a,b,c,d,e}$ are second-order transport coefficients. For the full expression of the second-order hydrodynamic equations including the scalar and the tensor dissipative currents, see for example Ref.~\cite{Monnai:2010qp}. 

The second law of thermodynamics implies that the entropy production is expressed in a quadratic form. It can be written in the Landau frame as
\begin{eqnarray}
\partial_\mu s^\mu = - \frac{V_L^\mu V^L_\mu }{\kappa_V} \geq 0,
\end{eqnarray}
and in the Eckart frame as
\begin{eqnarray}
\partial_\mu s^\mu = - \frac{W_E^\mu W^E_\mu}{\kappa_W} \geq 0,
\end{eqnarray}
with the mostly-minus metric. The first and the second order transport coefficients of the two frames are related by the identification of the entropy production:
\begin{eqnarray}
\kappa_V &=& \kappa_W \bigg( \frac{n}{e+P} \bigg)^2, \label{kappaLE}\\
\tau_V &=& \tau_W - \frac{\kappa_W}{(e+P)T}, \label{tauLE} \\
\chi_V^a &=& \chi_W^a - \frac{\tau_W nT}{e+P}, \\
\chi_V^b &=& \chi_W^b + \tau_W T - \frac{\kappa_W}{e+P}, \\
\chi_V^c &=&  \chi_W^c + \frac{\kappa_W}{(e+P)T}, \\
\chi_V^d &=& \chi_W^d + \frac{\kappa_W}{(e+P)T}, \\
\chi_V^e &=& \chi_W^e . \label{chieLE}
\end{eqnarray}
See Appendix \ref{sec:A} for the derivation. Those relations indicate that the full second-order terms are necessary in addition to the conventional Israel-Stewart terms for understanding the frame dependence of relativistic dissipative hydrodynamics, because the vanishing second-order transport coefficients in one frame lead to non-vanishing ones in the other frame, except for $\chi_V^e$ and $\chi_W^e$.

%%%%%%%%%%%%%%%%%%%%%%%%%%%%%%%%%%%%%%%%%%%%%%%%%%%%%%%%%%%%%%%%
\section{CAUSALITY AND STABILITY OF SECOND-ORDER HYDRODYNAMICS}
\label{sec3}
\vspace*{-2mm}
%%%%%%%%%%%%%%%%%%%%%%%%%%%%%%%%%%%%%%%%%%%%%%%%%%%%%%%%%%%%%%%%

In this section, causality and stability conditions of the relativistic full second-order hydrodynamic equations are investigated in the Landau and the Eckart frames. 
A plane wave perturbation $\delta Q = \delta \bar{Q} e^{i(\omega t - kx)}$ is considered for a macroscopic variable $Q$ around global equilibrium where $u^\mu = (1,0,0,0)$. The perturbed equations of motion are used to analyze the hydrodynamic modes \cite{Hiscock:1985zz,Hiscock:1987zz}.

\subsection{Landau Frame}

In the Landau frame, the perturbed energy-momentum tensor and the conserved charge current are 
\begin{eqnarray}
\delta T^{\mu \nu} &=& (e+P) (\delta u^\mu u^\nu + u^\mu \delta u^\nu) \nonumber \\
&+& \delta e u^\mu u^\nu - \delta P g^{\mu \nu}, \\
\delta N^\mu &=& n \delta u^\mu + \delta n u^\mu + \delta V^\mu ,
\end{eqnarray}
which follow the conservation law and the constitutive relation
\begin{eqnarray}
\delta V^\mu &=& \kappa_V \nabla^\mu \delta \alpha - \tau_V \Delta ^{\mu \nu} D \delta V_\nu ,
\end{eqnarray}
where $\alpha = \mu/T$. The longitudinal and the transverse modes relevant to the diffusion are given by
\begin{eqnarray}
\mathcal{M}^L_{xx}
\begin{pmatrix}
\delta e\\
\delta n\\
\delta u^x \\
\delta V^x \\
\end{pmatrix}
&=&0 , 
\end{eqnarray}
and
\begin{eqnarray}
\mathcal{M}^L_{xy}
\begin{pmatrix}
\delta u^y \\
\delta V^y \\
\end{pmatrix}
=0 , \
\mathcal{M}^L_{xz}
\begin{pmatrix}
\delta u^z \\
\delta V^z \\
\end{pmatrix}
=0 , 
\end{eqnarray}
where
\begin{eqnarray}
\mathcal{M}^L_{xx} = \begin{pmatrix}
i\omega&0&-ikh&0\\
-ik \frac{\partial P}{\partial e}|_n&-ik\frac{\partial P}{\partial n}|_e&i\omega h&0\\
0&i\omega&-ikn&-ik\\
-ik\kappa_V \frac{\partial \alpha}{\partial e}|_n&-ik\kappa_V \frac{\partial \alpha}{\partial n}|_e&0&1+i \omega \tau_V
\end{pmatrix} , \nonumber \\
\end{eqnarray}
and
\begin{eqnarray}
\mathcal{M}^L_{xy} = \mathcal{M}^L_{xz} =
\begin{pmatrix}
i\omega h &0 \\
0&1+i\omega \tau_V 
\end{pmatrix} ,
\end{eqnarray}
using the enthalpy density $h = e+P$. They have non-trivial solutions when the matrices have vanishing determinants. The longitudinal equations $\det(\mathcal{M}^L_{xx}) = 0$ lead to 
\begin{eqnarray}
\omega^2 - c_s^2 k^2 = \frac{i \kappa_V (c_2 \omega^2 - c_4 k^2) k^2 }{\omega (1+i \tau_V \omega)} , \label{eq:MLxx}
\end{eqnarray}
where the sound velocity is defined as 
\begin{eqnarray}
c_s^2 = \frac{\partial P}{\partial e}\bigg|_n + \frac{n}{h} \frac{\partial P}{\partial n}\bigg|_e , \label{eq:cs2}
\end{eqnarray}
and the thermodynamic coefficients as
\begin{eqnarray}
c_2 &=& \frac{\partial \alpha}{\partial n} \bigg|_e ,\\
c_4 &=& \frac{\partial \alpha}{\partial n} \bigg|_e \frac{\partial P}{\partial e}\bigg|_n  - \frac{\partial \alpha}{\partial e}\bigg|_n \frac{\partial P}{\partial n}\bigg|_e .
\end{eqnarray}
The Routh-Hurwitz stability criteria \cite{Routh,Hurwitz} indicate that the $\mathrm{Im} (\omega)$ stays semi-positive when $c_2 \geq 0$ and $c_s^2 c_2 - c_4 \geq 0$. Those conditions are satisfied in thermodynamic systems since the former follows from the thermodynamic requirement that the fugacity should increase as the number density increases at a fixed energy density and the latter from
\begin{eqnarray}
c_s^2 c_2 - c_4 &=& \frac{\beta}{h} 
\bigg( \frac{\partial P}{\partial \alpha}\bigg|_\beta \frac{\partial \alpha}{\partial n} \bigg|_e - \frac{\partial P}{\partial \beta}\bigg|_\beta \frac{\partial \alpha}{\partial e} \bigg|_n \bigg)^2 \geq 0 ,
\end{eqnarray}
where $\beta = 1/T$, using the definition of the sound velocity (\ref{eq:cs2}) and the thermodynamic properties
\begin{eqnarray}
\frac{\partial P}{\partial \alpha}\bigg|_\beta &=& \frac{n}{\beta} , \label{eq:dpda}\ \
\frac{\partial P}{\partial \beta}\bigg|_\alpha = - \frac{h}{\beta} ,
\end{eqnarray}
and
\begin{eqnarray}
\frac{\partial \beta}{\partial n} \bigg|_e &=& -  \frac{\partial \alpha}{\partial e} \bigg|_n \label{eq:bnae}.
\end{eqnarray}

Although it is possible to analytically solve the quartic equation, the general solutions are complicated. Here asymptotic forms at small $k$ are considered for more physical arguments. The propagating modes are, up to the leading order in real and imaginary parts,
\begin{eqnarray}
\omega = \pm c_s k + i \frac{\kappa_V (c_s^2 c_2 - c_4 )}{2 c_s^2} k^2 ,
\end{eqnarray}
and the non-propagating mode is
\begin{eqnarray}
\omega = \frac{i}{\tau_V} ,
\end{eqnarray}
aside from the trivial $\omega = 0$. They satisfy the causality condition
\begin{eqnarray}
\bigg | \frac{\partial \mathrm{Re} (\omega)}{\partial k} \bigg | \leq 1.
\end{eqnarray}
The stability condition 
\begin{eqnarray}
\mathrm{Im} (\omega) \geq 0,
\end{eqnarray}
is satisfied for $c_2 c_s^2 - c_4\geq 0$, which is consistent with the Routh-Hurwitz stability conditions.

The solutions to the transverse equations 
\begin{eqnarray}
\det(\mathcal{M}^L_{xy}) &=& \det(\mathcal{M}^L_{xz}) \nonumber \\
&=& i \omega h (1+i\omega \tau_V) = 0, \label{eq:MLxy}
\end{eqnarray}
are the non-propagating modes $\omega = 0$ and
$\omega = i/\tau_V$. The causality and stability conditions are trivially satisfied. Those results of the longitudinal and the transverse modes indicate that the second-order diffusive hydrodynamics is causal and stable in the Landau frame.

\subsection{Eckart Frame}

In the Eckart frame, the energy-momentum tensor and the conserved charge current are expressed as
\begin{eqnarray}
\delta T^{\mu \nu} &=& (e+P) (\delta u^\mu u^\nu + u^\mu \delta u^\nu) + \delta e u^\mu u^\nu \nonumber \\
&-& \delta P g^{\mu \nu} + \delta W^\mu u^\nu + \delta W^\nu u^\mu, \\
\delta N^\mu &=& n \delta u^\mu + \delta n u^\mu ,
\end{eqnarray}
and the energy dissipation current as
\begin{eqnarray}
\delta W^\mu &=& - \kappa_W \beta_\mathrm{eq} D \delta u^\mu - \kappa_W \nabla^\mu \delta \beta - \tau_W \Delta^{\mu \nu} D \delta W_\nu . \nonumber \\
\end{eqnarray}
The perturbed equations of motion are
\begin{eqnarray}
\mathcal{M}^E_{xx}
\begin{pmatrix}
\delta e\\
\delta n\\
\delta u^x \\
\delta W^x \\
\end{pmatrix}
&=&0 , 
\end{eqnarray}
and
\begin{eqnarray}
\mathcal{M}^E_{xy}
\begin{pmatrix}
\delta u^y \\
\delta W^y \\
\end{pmatrix}
=0 , \
\mathcal{M}^E_{xz}
\begin{pmatrix}
\delta u^z \\
\delta W^z \\
\end{pmatrix}
=0 , 
\end{eqnarray}
where
\begin{eqnarray}
\mathcal{M}^E_{xx} = \begin{pmatrix}
i\omega&0&-ik h&-ik\\
-ik \frac{\partial P}{\partial e}|_n&-ik \frac{\partial P}{\partial n}|_e&i\omega h&i\omega \\
0&i\omega&-ikn&0\\
ik\kappa_W \frac{\partial \beta}{\partial e}|_n&ik\kappa_W \frac{\partial \beta}{\partial n}|_e&i \omega \kappa_W \beta &1+i \omega \tau_W
\end{pmatrix} , \nonumber \\
\end{eqnarray}
and
\begin{eqnarray}
\mathcal{M}^E_{xy} = \mathcal{M}^E_{xz} =
\begin{pmatrix}
i\omega h&i \omega \\
i \omega \kappa_W \beta &1+i\omega \tau_V 
\end{pmatrix} .
\end{eqnarray}
The longitudinal equations $\det(\mathcal{M}^E_{xx}) = 0$ lead to 
\begin{eqnarray}
\omega^2 - c_s^2 k^2 = \frac{i \kappa_W (d_2 \omega^2 - d_4 k^2) k^2 }{\omega [1+i (\tau_W - \kappa_W \beta/h ) \omega]} , \label{eq:MExx}
\end{eqnarray}
where
\begin{eqnarray}
d_2 &=& \frac{n}{h} \bigg( \frac{\partial \beta}{\partial n} \bigg|_e + \frac{\beta}{h} \frac{\partial P}{\partial n}\bigg|_e \bigg),\\
d_4 &=& \frac{n}{h} \bigg( \frac{\partial \beta}{\partial n} \bigg|_e \frac{\partial P}{\partial e}\bigg|_n  - \frac{\partial \beta}{\partial e}\bigg|_n \frac{\partial P}{\partial n}\bigg|_e \bigg) .
\end{eqnarray}
Here, $\mathrm{Im} (\omega)$ stays semi-positive when 
$d_2 \geq 0$, $c_s^2 d_2  - d_4 \geq 0$, and
\begin{eqnarray}
\tau_W - \frac{\beta}{h} \kappa_W \geq 0, \label{eq:twkw}
\end{eqnarray}
according to the Routh-Hurwitz stability criteria. The first two conditions are again satisfied in thermodynamic systems as
\begin{eqnarray}
d_2 &=& \frac{n^2}{h^2} \frac{\partial \alpha}{\partial n}\bigg|_e \geq 0,\\
c_s^2 d_2  - d_4 &=& \frac{n^2 \beta}{h^3} 
\bigg( \frac{\partial P}{\partial \alpha}\bigg|_\beta \frac{\partial \alpha}{\partial n} \bigg|_e - \frac{\partial P}{\partial \beta}\bigg|_\beta \frac{\partial \alpha}{\partial e} \bigg|_n \bigg)^2 \geq 0,\nonumber \\
\end{eqnarray}
using the relations (\ref{eq:dpda}) and (\ref{eq:bnae}). Note that $d_2 = c_2 n^2/h^2$ and $d_4 = c_4 n^2/h^2$.
The third condition is also consistent with the ones reported in Ref.~\cite{Hiscock:1987zz,Osada:2011gx}.

The results indicate that second-order dissipative hydrodynamics is stable in the Eckart frame if the transport coefficients satisfy the condition (\ref{eq:twkw}). It can be immediately seen that the first-order theory is unstable in the Eckart frame by taking the limit of vanishing relaxation time $\tau_W\to 0$. 

It is important to note that the space-like projection of the energy-momentum conservation law leads to
\begin{eqnarray}
(e+P)Du^\mu &=& \nabla^\mu P
- W^\mu \nabla_\nu u^\nu \nonumber \\
&-& W^\nu \nabla_\nu u^\mu - \Delta^{\mu \nu} DW_\nu , \label{eq:emcperp}
\end{eqnarray}
which is also used to convert the thermodynamic forces (\ref{eq:thermoforce}).
The higher order terms in the identity is important even in the stability analyses of the first-order theory because if one neglects the correction by truncation and use it to remove the acceleration term in the energy dissipation current, the equation can become seemingly ``stable" at the first order. This is because the relaxation term-like correction originating from the last term in Eq.~(\ref{eq:emcperp}) is effectively introduced by the procedure at the second order even though it is not apparent. The prefactor before this effective relaxation term is $\kappa_W/(e+P)T$, which is the minimum value of the relaxation time required for hydrodynamic stability. The constitutive relation is qualitatively modified and thus cannot be regarded as a first-order theory. 

The asymptotic forms of the propagating and the non-propagating modes at small $k$ are 
\begin{eqnarray}
\omega = \pm c_s k + i \frac{\kappa_W (c_s^2 d_2 - d_4 )}{2 c_s^2} k^2 ,
\end{eqnarray}
and 
\begin{eqnarray}
\omega = \frac{i}{\tau_W - \kappa_W \beta/h} ,
\end{eqnarray}
aside from $\omega = 0$. Those modes are causal and stable if the Routh-Hurwitz criteria are satisfied.

The transverse equations 
\begin{eqnarray}
\det(\mathcal{M}^E_{xy}) &=& \det(\mathcal{M}^E_{xz}) \nonumber \\
&=& i\omega [h+i\omega (\tau_V - \kappa_W \beta)] = 0, \label{eq:MExy}
\end{eqnarray}
have the non-propagating solutions $\omega = i/(\tau_W - \kappa_W \beta/h)$ and $\omega = 0$. One can see that all the modes satisfy the stability and causality conditions if the relaxation time is sufficiently larger than the conductivity (\ref{eq:twkw}).

Comparing the two frames, the characteristic equations in the Landau frame (\ref{eq:MLxx}) and (\ref{eq:MLxy}) and their solutions are equivalent to those in the Eckart frame (\ref{eq:MExx}) and (\ref{eq:MExy}) under the identification of the conductivities (\ref{kappaLE}) and the relaxation times (\ref{tauLE}) that follow from the matching of the entropy production.
The relation of the relaxation times in the two frames implies that the Eckart stability condition on $\tau_W$ is closely related with the fact that $\tau_V$ is semi-positive in the other frame. 

%%%%%%%%%%%%%%%%%%%%%%%%%%%%%%%%%%%%%%%%%%%%%%%%%%%%%%%%%%%%%%%%
\section{Numerical Application to Heavy-Ion Collisions}
\label{sec4}
\vspace*{-2mm}
%%%%%%%%%%%%%%%%%%%%%%%%%%%%%%%%%%%%%%%%%%%%%%%%%%%%%%%%%%%%%%%%

The effects of a frame choice on relativistic nuclear collisions are demonstrated by solving the energy dissipative and the baryon diffusive hydrodynamic equations. For this purpose, a non-boost invariant (1+1)-dimensional hydrodynamic system is considered \cite{Monnai:2012jc}. Full (3+1)-dimensional calculations for quantitative analyses of the data sets from the beam energy scan experiments is beyond the scope of the current study and will be presented elsewhere.

\subsection{The hydrodynamic model}

Hydrodynamics system are characterized with the equation of state and the transport coefficients. 
The equation of state at finite baryon density \cite{Monnai:2019hkn} is based on lattice QCD \cite{Bazavov:2014pvz,Bazavov:2012jq,Ding:2015fca,Bazavov:2017dus} and the hadron resonance gas model. The strangeness and the electric charge are not considered here for simplicity and left for future studies.

The transport coefficients are chosen as $\kappa_W = c_W (e+P)$, 
$\tau_W = \tilde{c}_W \kappa_W / (e+P)T$, and $\chi_W^{a,b,c,d,e} = 0$ in the Eckart frame. The model conductivity is motivated by the non-equilibrium statistical operator method for the $\phi^4$-theory \cite{Hosoya:1983id} coupled with the lower bound of shear viscosity conjectured in the gauge-string correspondence \cite{Kovtun:2004de}. $c_W = 10$ and $\tilde{c}_W = 2$ are used for demonstration. Those in the Landau frame is obtained using the relations (\ref{kappaLE})-(\ref{chieLE}). 

The initial conditions are parametrically constructed as
\begin{eqnarray}
e(\tau_\mathrm{th},\eta_s) &=& a_1 \exp(-a_2 \eta_s^2 - a_3 \eta_s^4), \\
n_B(\tau_\mathrm{th},\eta_s) &=& n_B^+(\eta_s) + n_B^-(\eta_s),
\end{eqnarray}
where
\begin{eqnarray}
&&n_B^\pm(\eta_s) = \nonumber \\
&&\begin{cases}
b_1 \exp[-b_2 (\eta_s \mp \eta_0)^2 - b_3 (\eta_s \mp \eta_0)^4] & \mathrm{for} \ \pm \eta_s > \eta_0, \\
b_1 \exp[-\tilde{b}_2 (\eta_s \mp \eta_0)^2 - \tilde{b}_3 (\eta_s \mp \eta_0)^4] & \mathrm{for} \ \pm \eta_s \leq \eta_0,
\end{cases}\nonumber \\
\end{eqnarray}
at $\tau_\mathrm{th} = 3$ fm/$s$. The parameters are tuned to roughly reproduce the SPS data for 17.3 GeV Pb+Pb collisions \cite{Appelshauser:1998yb} without dissipative corrections. Here $a_1= 7.19$~(GeV/fm$^{3}$), $a_2 = 0.8$, and $a_3 = 0.05$ for the energy density and $b_1 = 0.45$~(1/fm$^{3}$), $b_2 = 0.4$, $b_3 = 4.0$, $\tilde{b}_2 = 0.55$, $\tilde{b}_3 = 2.3$, and $\eta_0 = 0.69$ for the net baryon density. It should be noted again that they are for demonstration and not for full quantitative analyses of the data because even though the results exhibit fair agreement with the data, the transverse expansion and the hadronic transport are not taken into account here. The prolonged space-time evolution may partially mimic the transport effects. The initial values of the energy dissipation and the baryon diffusion currents are set to zero to allow comparison of the effects of those processes coming from hydrodynamic evolution. 

The kinetic freeze-out is estimated using the Cooper-Frye formula \cite{Cooper:1974mv} with off-equilibrium corrections to the phase-space distribution functions \cite{Teaney:2003kp,Monnai:2009ad}. It reads
\begin{eqnarray}
E_i\frac{dN^i}{d^3p} = \frac{g_i}{(2\pi)^3} \int_\Sigma p_i^\mu d\sigma_\mu (f^0_i + \delta f_i), \label{eq:CF}
\end{eqnarray}
where $g_i$ is the degeneracy, $\Sigma$ is the freeze-out hypersurface, and $d\sigma_\mu$ is the freeze-out hypersurface element. $f_i^0$ is the equilibrium (Bose-Einstein or Fermi-Dirac) phase-space distribution function for the $i$-th particle species and $\delta f_i$ is the off-equilibrium distortion of the distribution function. The expression of $\delta f$ in the Landau and the Eckart frames are shown in Appendix~\ref{sec:B}. The hypersurface is determined with the freeze-out energy density $e_\mathrm{f} = 0.4$~GeV/fm$^3$. 

\subsection{Space-time evolution}
\label{sec4B}

%%%%%%%%%%%%%%%%%%%%%%%%%%%%%%%%%%%%%%%%%%%%%%%%%%%%%%%%%%%%%%%%
\begin{figure}[tb]
\includegraphics[width=3.3in]{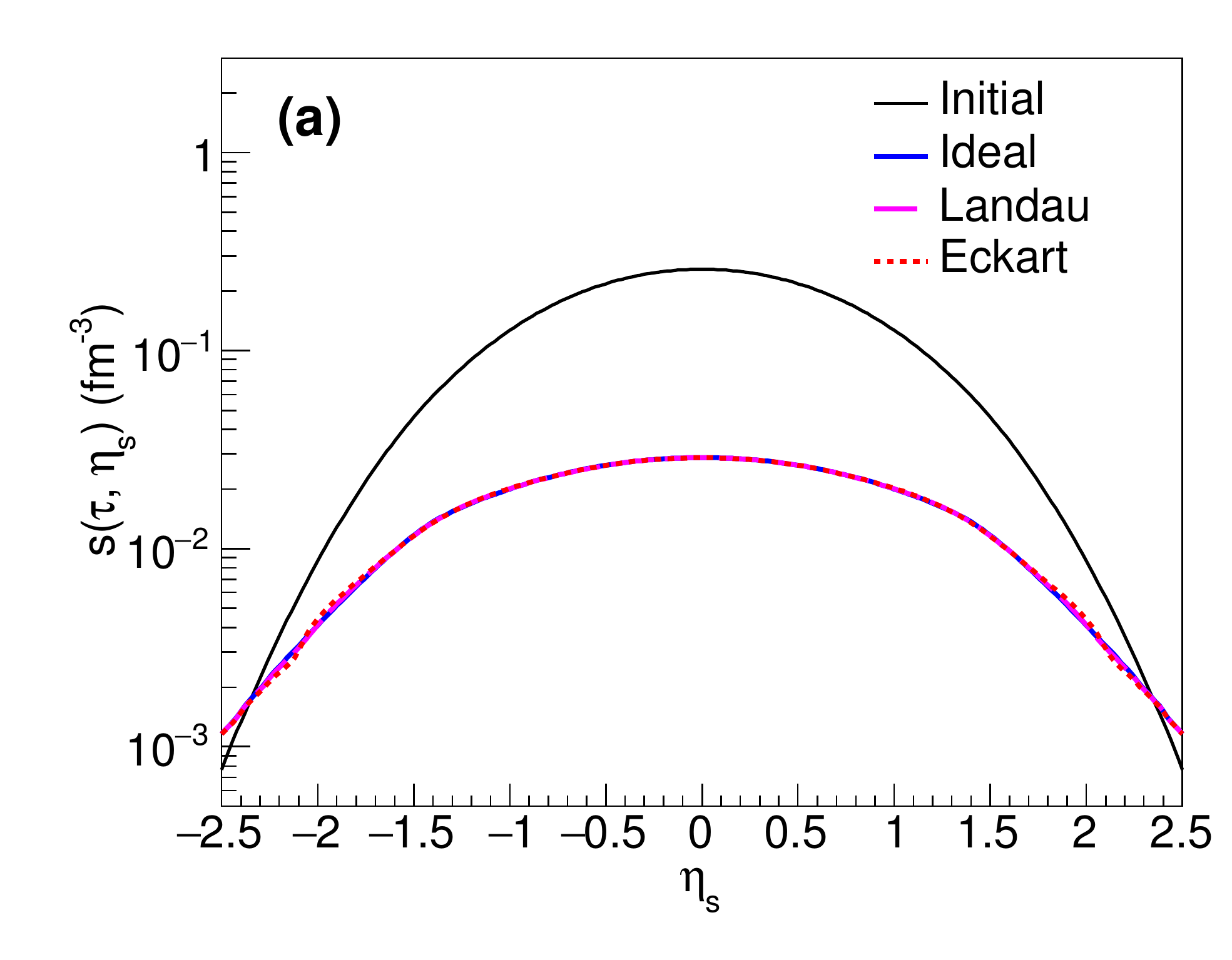}
\includegraphics[width=3.3in]{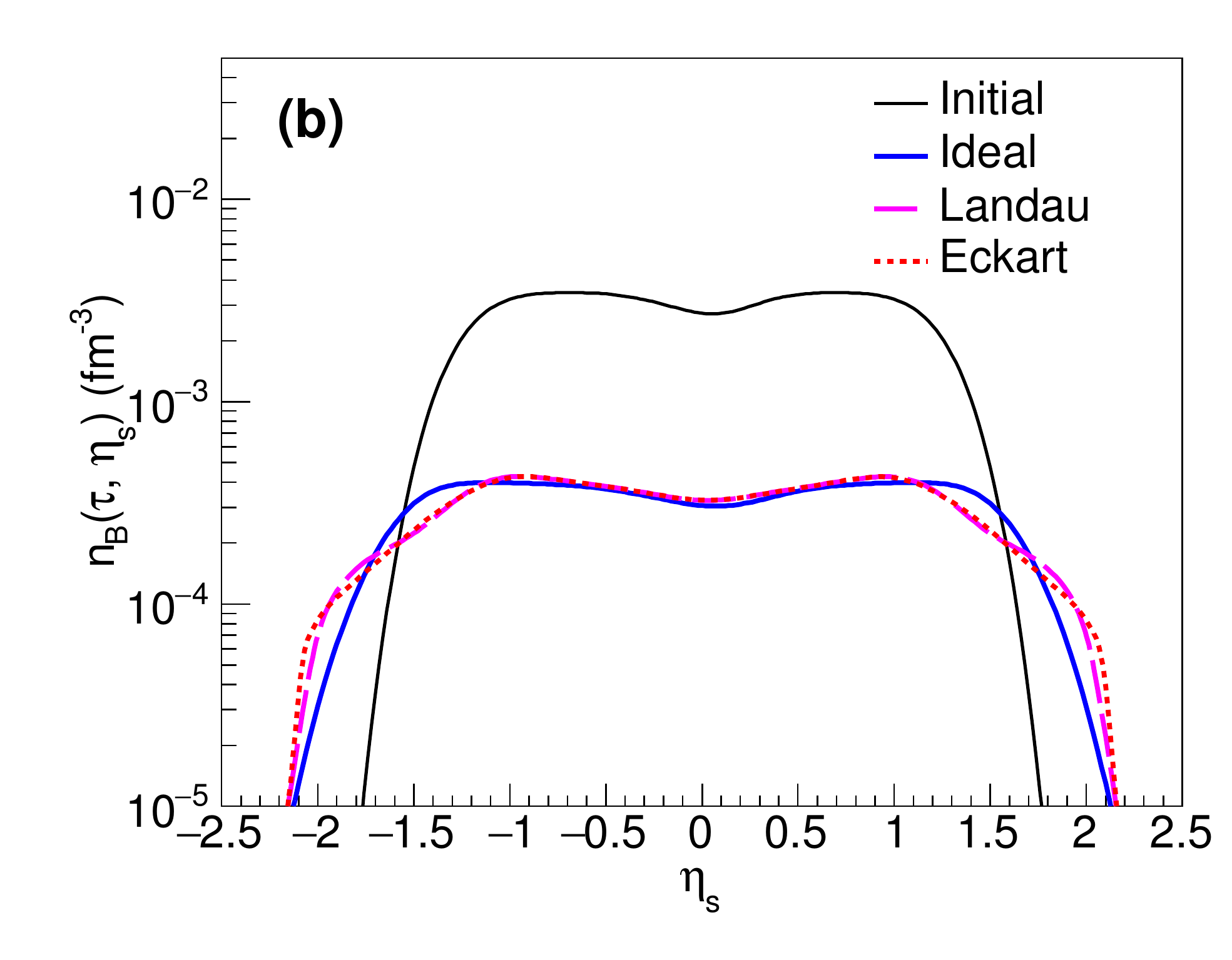}
\caption{The space-time rapidity dependences of (a) the entropy density and (b) the net baryon density at the initial time (thin solid line) and those after ideal (thick solid line), baryon diffusive (dashed line), and energy dissipative (dotted line) hydrodynamic evolutions at $\tau = 20$ fm/$c$.}
\label{fig:1}
\end{figure}
%%%%%%%%%%%%%%%%%%%%%%%%%%%%%%%%%%%%%%%%%%%%%%%%%%%%%%%%%%%%%%%%

First, I investigate the off-equilibrium hydrodynamic evolution in the Landau and the Eckart frames and compare them with the ideal hydrodynamic evolution. The entropy and the net baryon distributions at the initial time and $\tau = 20$~fm/$c$ are shown in Fig.~\ref{fig:1}. It should be noted that the lifetime of the fireball is longer in the current geometry owing to the lack of transverse expansion. The effect of baryon diffusion or energy dissipation is small on the entropy density for the current choice of transport coefficients. 

\begin{figure}[tb]
\includegraphics[width=3.3in]{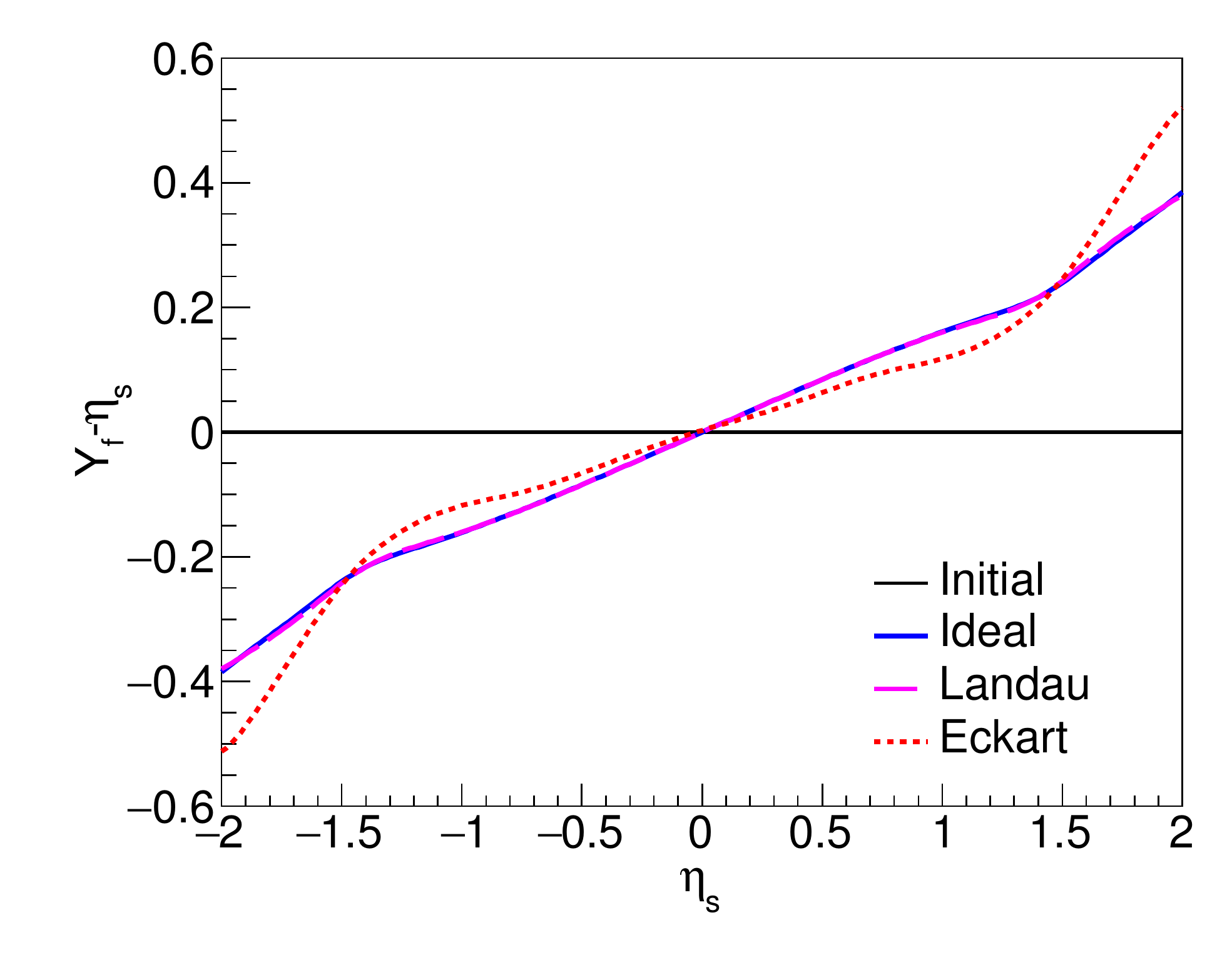}
\caption{The space-time rapidity dependence of the difference between the flow and the space-time rapidities at the initial time (thin solid line) and those after ideal (thick solid line), baryon diffusive (dashed line), and energy dissipative (dotted line) hydrodynamic evolutions at $\tau = 20$ fm/$c$.}
\label{fig:2}
\end{figure}

The effect on the net baryon density, on the other hand, is visible. The baryon diffusion causes stronger stopping because the fugacity gradients induce net baryon diffusion from forward to mid-rapidity regions. At the edges near $|\eta_s| \sim 2$, the baryon diffusion is in the outward direction. The energy dissipation, on the other hand, is less trivial because of the interplay of the temperature gradient and the acceleration terms. The temperature gradients carry the energy density towards forward rapidity regions while the acceleration correction prevents flow convection and keep the density in the mid-rapidity region. The effects cancel at the first-order in the limit of vanishing chemical potential as seen in (\ref{eq:thermoforce}). The off-equilibrium deformation of the net baryon distribution in the Eckart frame can be mainly caused by the deceleration of flow as seen in Fig.~\ref{fig:2} near mid-rapidity. The off-equilibrium evolutions of the net baryon distribution in the Landau frame and in the Eckart frame are quantitatively similar to each other. This can be a consequence of the fact that the frame-dependence of the thermodynamic quantities are of second order.

The difference between the flow rapidity $Y_f$ and the space-time rapidity $\eta_s$ (Fig.~\ref{fig:2}) implies that the Landau flow is closer to the ideal flow than the Eckart flow. Here the flow rapidity is defined as 
\begin{eqnarray}
u^\mu = (\cosh Y_f,0,0,\sinh Y_f),
\end{eqnarray}
which reduces to the boost-invariant flow when $Y_f - \eta_s = 0$. The flow is affected more in the Eckart frame possibly because the energy dissipation is directly coupled to the equation of motion for flow acceleration (\ref{eq:emcperp}). At forward space-time rapidity $|\eta_s| > 1.5$, the Eckart flow is faster then the Landau flow because of the peak position in the net baryon distribution.

%%%%%%%%%%%%%%%%%%%%%%%%%%%%%%%%%%%%%%%%%%%%%%%%%%%%%%%%%%%%%%%%
%%%%%%%%%%%%%%%%%%%%%%%%%%%%%%%%%%%%%%%%%%%%%%%%%%%%%%%%%%%%%%%%

\subsection{Charged particle and net baryon rapidity distributions}

\begin{figure}[tb]
\includegraphics[width=3.3in]{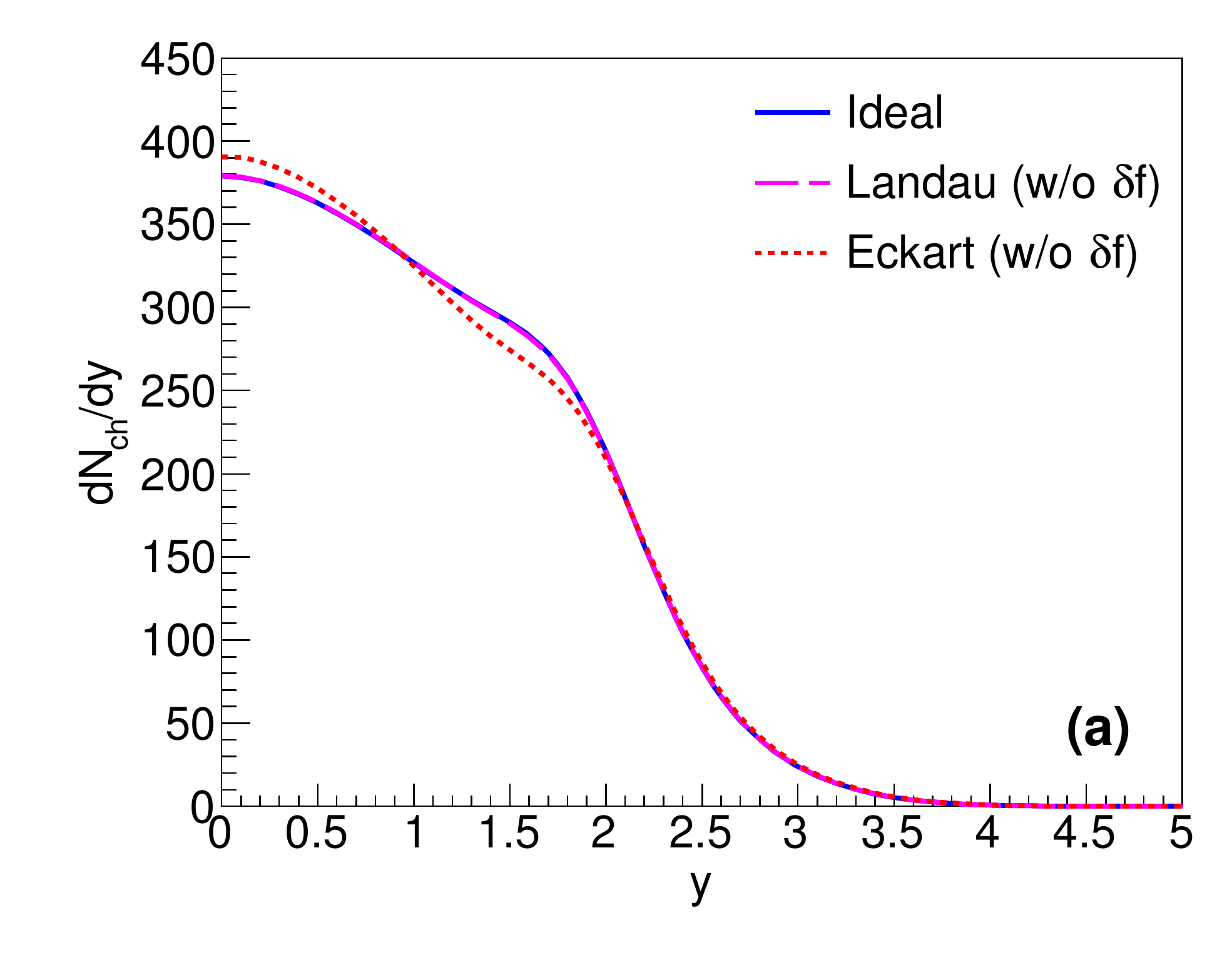}
\includegraphics[width=3.3in]{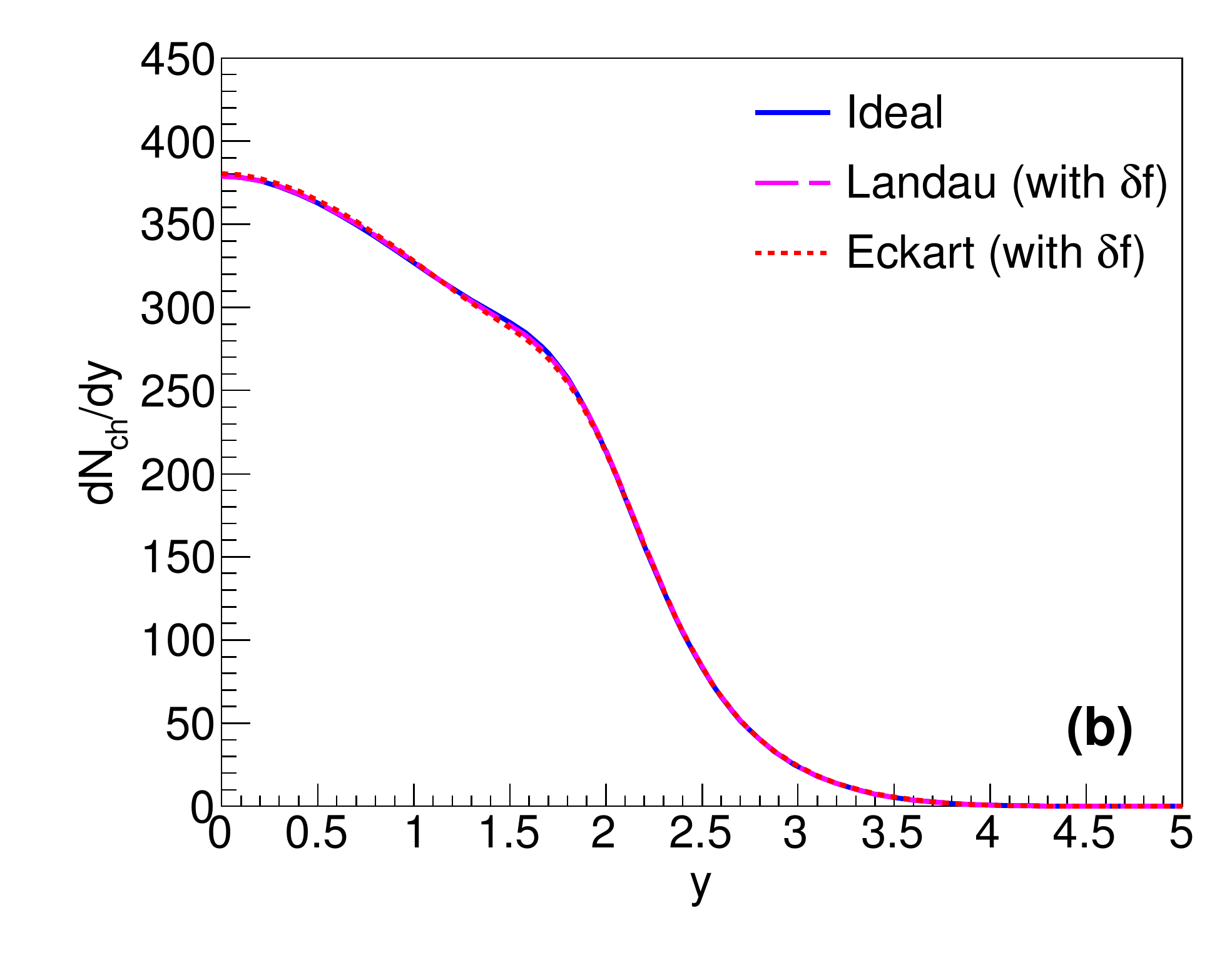}
\caption{The rapidity distributions of charged particles (a) without and (b) with $\delta f$ correction at freeze-out for the ideal hydrodynamic system (solid line) compared to those for the systems with baryon diffusion in the Landau frame (dashed line) and with energy dissipation in the Eckart frame (dotted line).}
\label{fig:3}
\end{figure}

The charged hadron rapidity distributions are shown in Fig.~\ref{fig:3}. The effect of energy dissipation in the Eckart frame is visible while that of baryon diffusion is negligible when the off-equilibrium correction at freeze-out  (\ref{eq:CF}) is not taken into account. The difference comes from the difference in the Landau and the Eckart flow and the lack of the $\delta f$ corrections.
When the correction is incorporated, the effect of energy dissipation becomes small and similar to that of baryon diffusion as found in Fig.~\ref{fig:3} (b). 

%%%%%%%%%%%%%%%%%%%%%%%%%%%%%%%%%%%%%%%%%%%%%%%%%%%%%%%%%%%%%%%%
%%%%%%%%%%%%%%%%%%%%%%%%%%%%%%%%%%%%%%%%%%%%%%%%%%%%%%%%%%%%%%%%

\begin{figure}[tb]
\includegraphics[width=3.3in]{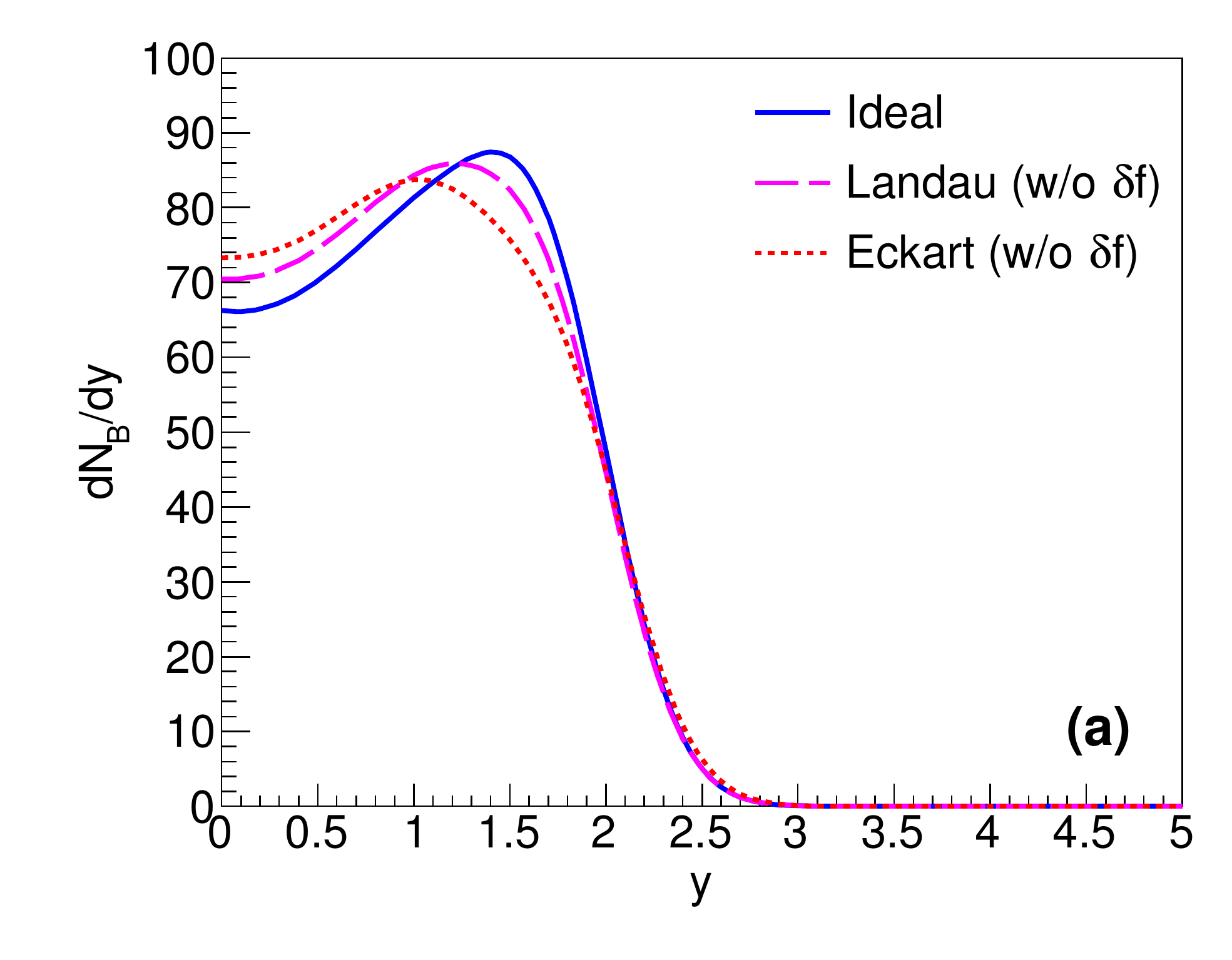}
\includegraphics[width=3.3in]{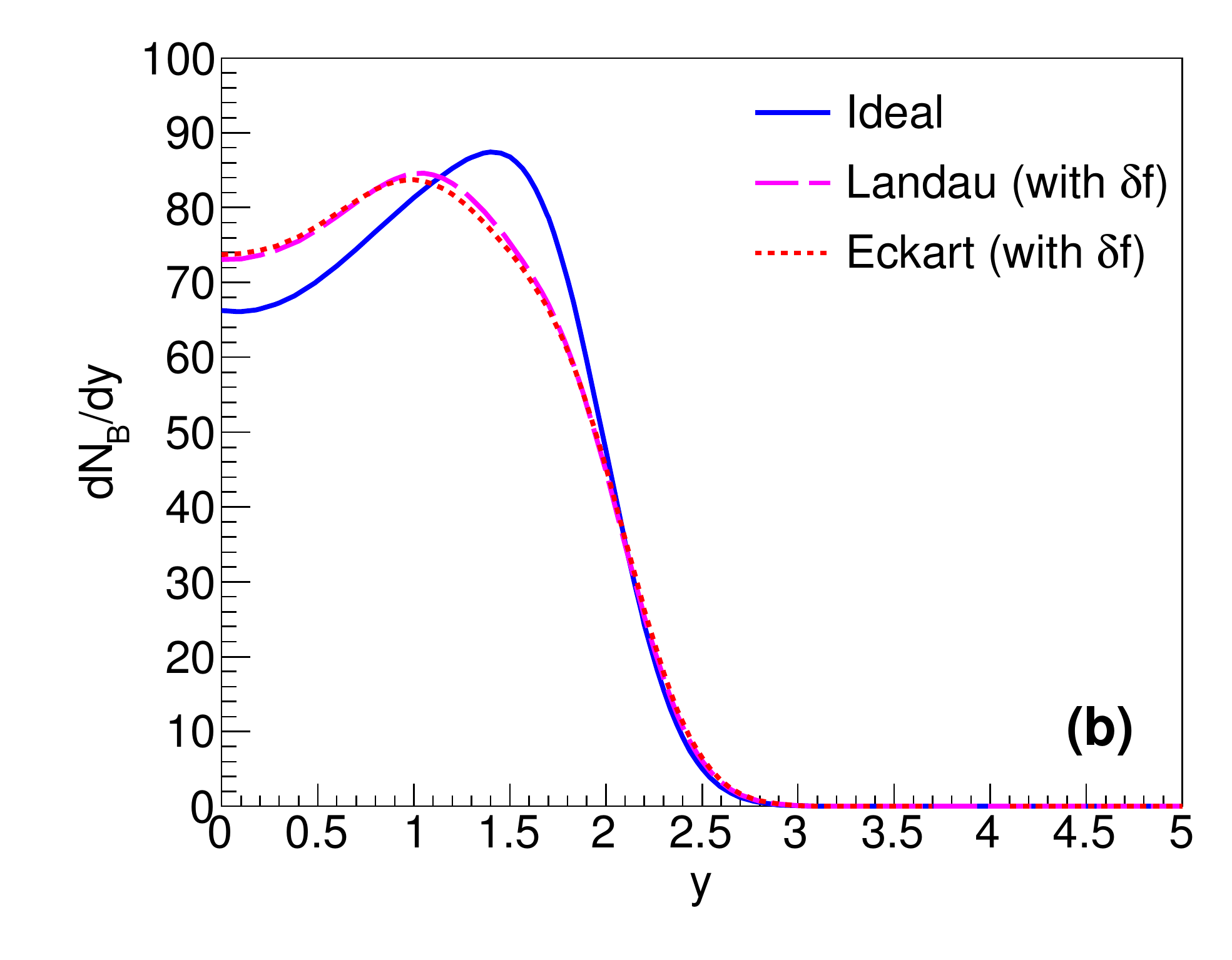}
\caption{The rapidity distributions of net baryon number (a) without and (b) with $\delta f$ correction at freeze-out for the ideal hydrodynamic system (solid line) compared to those for the systems with baryon diffusion in the Landau frame (dashed line) and with energy dissipation in the Eckart frame (dotted line).}
\label{fig:4}
\end{figure}

The net baryon rapidity distribution with baryon diffusion in the Landau frame and with energy dissipation in the Eckart frame are shown in Fig.~\ref{fig:4}. 
The off-equilibrium effects are visible in both frames without the $\delta f$ correction. This is consistent with the observation of hydrodynamic evolution of the net baryon density in Sec.~\ref{sec4B}. The baryon stopping is larger in the Eckart frame because of the flow deceleration.
The effect of $\delta f$ correction at freeze-out is found to enhance the baryon stopping caused by the baryon diffusion. Again the net baryon distributions in the two frames become close to each other once the off-equilibrium correction at freeze-out is properly taken into account.

%%%%%%%%%%%%%%%%%%%%%%%%%%%%%%%%%%%%%%%%%%%%%%%%%%%%%%%%%%%%%%%%
%%%%%%%%%%%%%%%%%%%%%%%%%%%%%%%%%%%%%%%%%%%%%%%%%%%%%%%%%%%%%%%%

\begin{figure}[tb]
\includegraphics[width=3.2in]{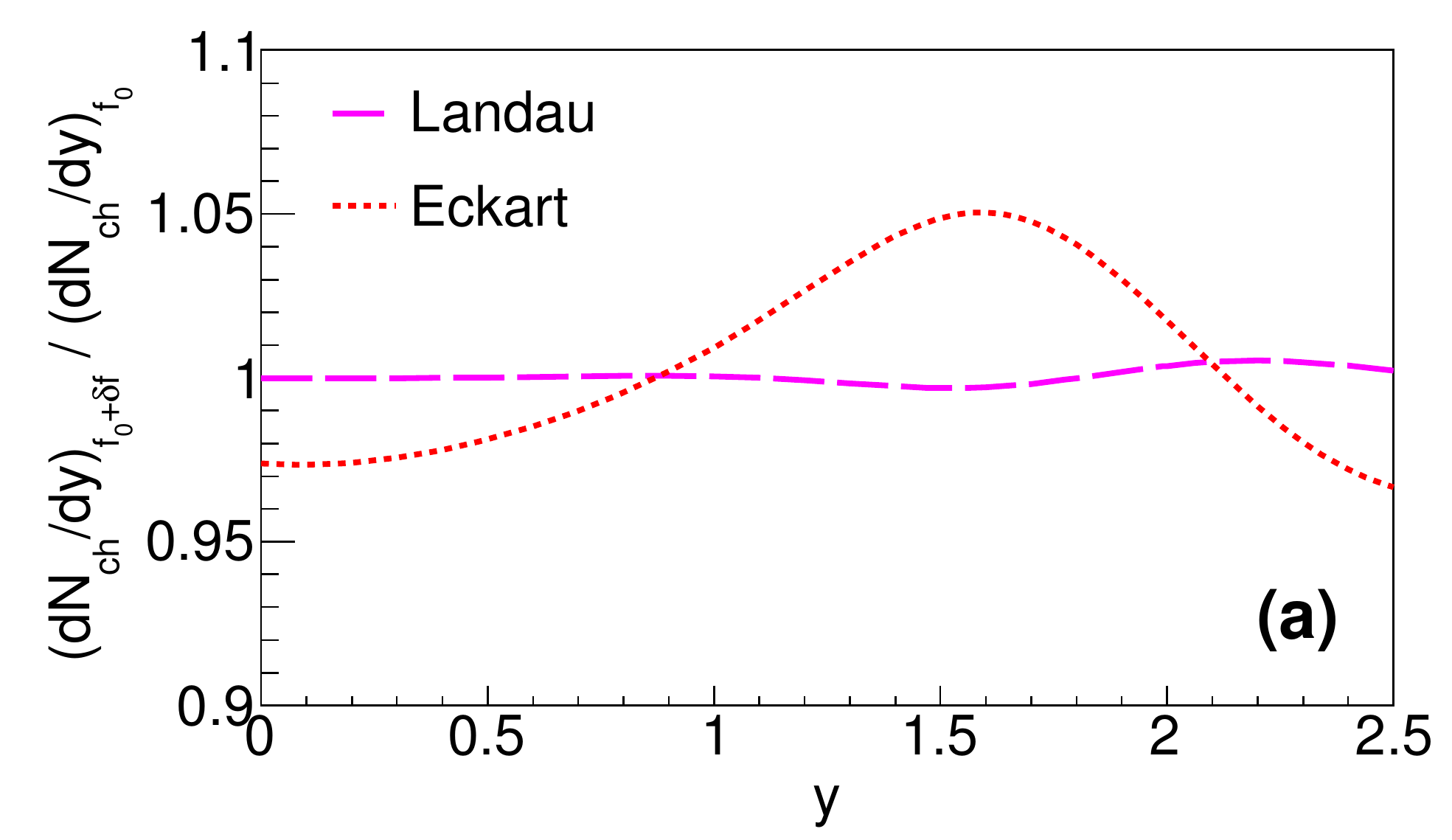}
\includegraphics[width=3.2in]{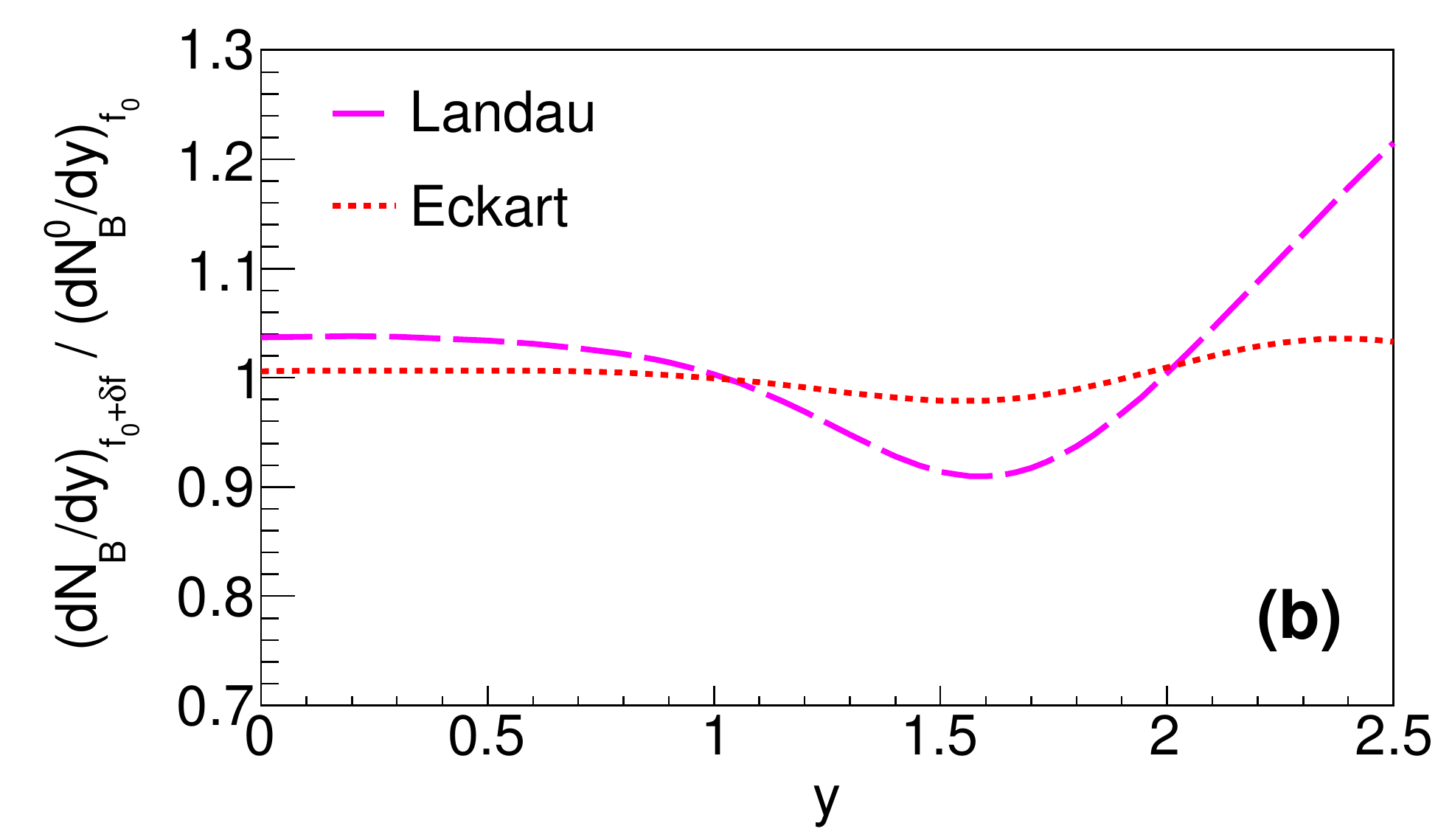}
\caption{The ratios of the rapidity distributions with and without $\delta f$ correction for (a) charged particles and (b) for net baryon number in the Landau frame (dashed line) and in the Eckart frame (dotted line).}
\label{fig:5}
\end{figure}

It is worth noting that the effect of the $\delta f$ correction is larger in the Eckart frame for the charged particle distribution while it is larger in the Landau frame for the net baryon distribution (Fig.~\ref{fig:5}). The results suggest that an adequate treatment of $\delta f$ corrections are important for qualitative understanding of the flow observables.

%%%%%%%%%%%%%%%%%%%%%%%%%%%%%%%%%%%%%%%%%%%%%%%%%%%%%%%%%%%%%%%%
%%%%%%%%%%%%%%%%%%%%%%%%%%%%%%%%%%%%%%%%%%%%%%%%%%%%%%%%%%%%%%%%

%%%%%%%%%%%%%%%%%%%%%%%%%%%%%%%%%%%%%%%%%%%%%%%%%%%%%%%%%%%%%%%%
\section{Discussion and Conclusions}
\label{sec5}
\vspace*{-2mm}
%%%%%%%%%%%%%%%%%%%%%%%%%%%%%%%%%%%%%%%%%%%%%%%%%%%%%%%%%%%%%%%%

The baryon diffusive and the energy dissipative hydrodynamics at the second-order in the Landau and the Eckart frames have been discussed. The system is stable at the second order when the relaxation time is semi-positive in the Landau frame and it is larger than the minimum value in the Eckart frame. The mode analyses implies that causality is also satisfied in the long wave length limit. The transport coefficients of the two frames at the linear and the second order are shown to be related. The full second-order terms are found to be necessary for a consistent matching. The results are generic and independent of the individual derivation method of the hydrodynamic equations of motion.

The frame dependence is tested in a numerical hydrodynamic model of relativistic heavy-ion collisions. The net baryon number is chosen as the conserved charge of the system and the space-time evolutions of a QCD medium in the Landau and the Eckart frames are compared to that of the inviscid system. The space-time rapidity distribution of the entropy density is not much affected by the dissipative currents while that of net baryon density is visibly modified. The effects of the baryon diffusion and the energy dissipation is found to be quantitatively similar for those thermodynamic variables. The flow, on the other hand, is implied to be different in the Landau and the Eckart frames.

The charged particle distribution is estimated in both frames. The result is found to be mostly unaffected by the baryon diffusion for the chosen set of transport coefficients. The distribution for the energy dissipation is also not modified much owing to the cancellation of the effects of the flow deceleration and the off-equilibrium correction at freeze-out. A larger baryon stopping is observed in the net baryon distribution owing to the fugacity gradient for the Landau frame and also to the flow deceleration in the Eckart frame. The $\delta f$ correction is found to increase the baryon stopping effect of baryon diffusion so that the difference between the net baryon distributions of the two frames becomes small.

The results indicate that the hydrodynamic estimation of the observables may not depend much on the choice of the local rest frame in relativistic nuclear collisions. It would be important to investigate other observables that are directly dependent on the flow, such as thermal photons with blue shifting, to elucidate the issue of the Landau and the Eckart frames in hydrodynamic models.

It is worth noting that studied in the present numerical analyses are the finite temperature and chemical potential regions near the QCD transition explored by relativistic nuclear collisions. One should be careful when determining a frame in the zero temperature or chemical potential limit. A careful treatment of the equation of state and the transport coefficients may also become important in such cases.

Future prospects include the application to the full (3+1) dimensional analyses of the beam energy scan data of flow-related observables to extract relations between the initial conditions and the transport coefficients in each frame to investigate the validity of the choice of the local rest frame more quantitatively. 

\begin{acknowledgments}
The author is grateful for the valuable comments by T.~Kunihiro. 
The work of A.M. was supported by JSPS KAKENHI Grant Number JP19K14722.
\end{acknowledgments}

\appendix
%%%%%%%%%%%%%%%%%%%%%%%%%%%%%%%%%%%%%%%%%%%%%%%%%%%%%%%%%%%%%%%%
\section{ENTROPY PRODUCTION IN LANDAU AND ECKART FRAMES}
\label{sec:A}
\vspace*{-2mm}
%%%%%%%%%%%%%%%%%%%%%%%%%%%%%%%%%%%%%%%%%%%%%%%%%%%%%%%%%%%%%%%%

The relation between the transport coefficients can be determined by the identification of the entropy production of the Landau and the Eckart frames:
\begin{eqnarray}
\partial_\mu s^\mu = - \frac{V_L^\mu V^L_\mu}{\kappa_V} = - \frac{W_E^\mu W^E_\mu}{\kappa_W} .
\end{eqnarray}
The entropy production in the Landau frame up to the next-to-leading order is
\begin{eqnarray}
\partial_\mu s^\mu &=& - \kappa_V \nabla_\mu^L \frac{\mu}{T} \nabla^\mu_L \frac{\mu}{T} \nonumber \\
&+& 2 \tau_V \nabla_\mu^L \frac{\mu}{T} D_L V_L^\mu - 2 \chi_W^a \nabla_\mu^L \frac{\mu}{T} V_L^\mu D_L \frac{\mu}{T} \nonumber \\
&-& 2 \chi_V^b \nabla_\mu^L \frac{\mu}{T} V_L^\mu D_L \frac{1}{T} - 2 \chi_V^c \nabla_\mu^L \frac{\mu}{T} V_L^\mu \nabla^L_\nu u_L^\nu \nonumber \\
&-& 2 \chi_W^d \nabla_\mu^L \frac{\mu}{T} V_L^\nu \nabla^L_\nu u_L^\mu - 2 \chi_W^e \nabla_\mu^L \frac{\mu}{T} V_L^\nu \nabla_L^\mu u^L_\nu + \mathcal{O}(\delta^4). \nonumber \\
\end{eqnarray}

The entropy production in the Eckart frame can be expressed using the variables in the Landau frame as, up to the same order,
\begin{eqnarray}
\partial_\mu s^\mu &=& - \frac{W_E^\mu W^E_\mu}{\kappa_W} \nonumber \\
&=& - \kappa_W \bigg( \frac{n}{e+P} \bigg)^2 \nabla_\mu^L \frac{\mu}{T} \nabla^\mu_L \frac{\mu}{T} \nonumber \\
&+& 2 \bigg[ \tau_W - \frac{\kappa_W}{(e+P)T} \bigg] \nabla_\mu^L \frac{\mu}{T} D_L V_L^\mu \nonumber \\
&-& 2 \bigg[ \chi_W^a - \frac{\tau_W nT}{e+P} \bigg] \nabla_\mu^L \frac{\mu}{T} V_L^\mu D_L \frac{\mu}{T} \nonumber \\
&-& 2 \bigg[ \chi_W^b + \tau_W T - \frac{\kappa_W}{(e+P)} \bigg] \nabla_\mu^L \frac{\mu}{T} V_L^\mu D_L \frac{1}{T} \nonumber \\
&-& 2 \bigg[ \chi_W^c + \frac{\kappa_W}{(e+P)T} \bigg] \nabla_\mu^L \frac{\mu}{T} V_L^\mu \nabla^L_\nu u_L^\nu \nonumber \\
&-& 2 \bigg[ \chi_W^d + \frac{\kappa_W}{(e+P)T} \bigg] \nabla_\mu^L \frac{\mu}{T} V_L^\nu \nabla^L_\nu u_L^\mu \nonumber \\
&-& 2 \chi_W^e \nabla_\mu^L \frac{\mu}{T} V_L^\nu \nabla_L^\mu u^L_\nu + \mathcal{O}(\delta^4).
\end{eqnarray}
It should be noted that the thermodynamic forces of the energy dissipation and the baryon diffusion are mutually convertible using the hydrodynamic identity derived from the Gibbs-Duhem relation and energy-momentum conservation as
\begin{align}
&\bigg( \nabla_E^\mu \frac{1}{T} + \frac{1}{T} D_E u^\mu \bigg) = \frac{n}{e+P} \nabla_E^\mu \frac{\mu}{T} \nonumber \\
&- \frac{1}{(e+P)T} [W_E^\mu \nabla^E_\nu u_E^\nu + W_E^\nu \nabla^E_\nu u_E^\mu + (\Delta_{E})^{\mu}_{\ \nu} D_E W_E^\nu] \nonumber \\
&= \frac{n}{e+P} \nabla_L^\mu \frac{\mu}{T} - \frac{n}{e+P} \bigg( u_L^\mu \frac{V_L^\nu}{n} \nabla^L_\nu \frac{\mu}{T}  + \frac{V_L^\mu}{n} D^L \frac{\mu}{T}  \bigg) \nonumber \\
&+ \frac{1}{nT} \bigg[V_L^\mu \nabla^L_\nu u_L^\nu + V_L^\nu \nabla^L_\nu u_L^\mu + (\Delta_{L})^{\mu}_{\ \nu} D_L V_L^\nu \nonumber \\
&+ \frac{n}{e+P} V_L^\mu D_L \frac{e+P}{n} \bigg] + \mathcal{O}(\delta^3) , \label{eq:thermoforce}
\end{align}
where
\begin{eqnarray}
D_L \frac{e+P}{n} &=& - \frac{e+P}{n} T D_L \frac{1}{T} +T D_L \frac{\mu}{T} .
\end{eqnarray}
The correspondences between the transport coefficients in the two frames can be obtained as Eqs.~(\ref{kappaLE})-(\ref{chieLE}).

%%%%%%%%%%%%%%%%%%%%%%%%%%%%%%%%%%%%%%%%%%%%%%%%%%%%%%%%%%%%%%%%
\section{FREEZE-OUT WITH OFF-EQUILIBRIUM DISTRIBUTION}
\label{sec:B}
\vspace*{-2mm}
%%%%%%%%%%%%%%%%%%%%%%%%%%%%%%%%%%%%%%%%%%%%%%%%%%%%%%%%%%%%%%%%

The distribution function in relativistic systems with the energy dissipation and the baryon diffusion is estimated using the Grad's moment method \cite{Grad} based on Ref.~\cite{Israel:1979wp,Monnai:2010qp}. The distribution can be decomposed into the equilibrium and the off-equilibrium parts as
\begin{eqnarray}
f_0^i &=& \{\exp[(p^\mu u_\mu - b_i \mu_B)/T]\mp 1\}^{-1}, \\
\delta f^i &=& -f_0^i (1\pm f_0^i) (b_i p_i^\mu \varepsilon_\mu^B + p_i^\mu p_i^\nu \varepsilon_{\mu\nu}),
\end{eqnarray}
where $b_i$ is the quantum number for baryons. The upper sign is for bosons and the lower one for fermions. If the auxiliary vector and tensor $\varepsilon_\mu^B$ and $\varepsilon_{\mu \nu}$ are expressed in terms of macroscopic dissipative currents,
\begin{eqnarray}
\varepsilon_\mu^{L;B} = D_V V^L_\mu, \ \ \varepsilon^L_{\mu \nu} = B_V (V^L_\mu u^L_\nu + V^L_\nu u^L_\mu),
\end{eqnarray}
in the Landau frame and 
\begin{eqnarray}
\varepsilon_\mu^{E;B} = D_W W^E_\mu, \ \ \varepsilon^E_{\mu \nu} = B_W (W^E_\mu u^E_\nu + W^E_\nu u^E_\mu),
\end{eqnarray}
in the Eckart frame. The coefficients can be determined by the self-consistency condition that the off-equilibrium distribution reproduces the respective dissipative current within the framework of kinetic theory. They are,
\begin{eqnarray}
D_W = - 2J_{31}^B \mathcal{J}_2^{-1}, \ \ B_W = J_{21}^{BB} \mathcal{J}_2^{-1},
\end{eqnarray}
and 
\begin{eqnarray}
D_V = 2J_{41} \mathcal{J}_2^{-1}, \ \ B_V = -J_{31}^B \mathcal{J}_2^{-1},
\end{eqnarray}
where 
\begin{eqnarray}
\mathcal{J}_2 = 2(J_{31}^B J_{31}^B-J_{41}J_{21}^{BB}).
\end{eqnarray}
Here the moments are defined as
\begin{eqnarray}
J^{B...B}_{k l} &=& \frac{1}{(2l+1)!!} \sum_i \int \frac{(b_i ... b_i) d^3p}{(2\pi)^3 E_i} \nonumber \\
&\times& [m_i^2 - (p\cdot u)^2]^{l} (p\cdot u)^{k-2l} f_0^i(1\pm f_0^i) .
\end{eqnarray}

The off-equilibrium corrections are essential for conserving the energy-momentum and the net baryon number during the conversion from fluid to particles at freeze-out. The underlying equations of state for the hydrodynamic model and relativistic kinetic theory should be the same for successful conversion. The hadron gas with all resonances below 2 GeV in mass \cite{Tanabashi:2018oca} is used for the numerical estimation of the distortion coefficients to match the constructions of $\delta f$ and the equation of state.

\bibliography{frame.bib}

\end{document}